\documentclass{aa}
\usepackage{graphicx}

\begin{document}


%

   \title{BeppoSAX observations of Mrk~841  and Mrk~335}

   \author{S. Bianchi
          \inst{1}
    \and G. Matt\
          \inst{1}
\and F. Haardt
          \inst{2}
\and L. Maraschi
          \inst{3}
\and F. Nicastro
          \inst{4}
\and G.C. Perola
          \inst{1}
\and P.O. Petrucci 
          \inst{3}
\and L. Piro
          \inst{5}
          }

   \offprints{S. Bianchi, bianchi@fis.uniroma3.it}

   \institute{Dipartimento di Fisica, Universit\`a degli Studi ``Roma Tre",
              Via della Vasca Navale 84, I--00146 Roma, Italy
   \and Dipartimento di Scienze, Universit\`a dell'Insubria, Via Lucini 3,
         I--22100 Como, Italy
   \and Osservatorio Astronomico di Brera, Via Brera 28, I--20121 Milano, Italy
     \and Harvard-Smithsonian Center for Astrophysics,
                60 Garden st., Cambridge MA 02138 USA
  \and Istituto di Astrofisica Spaziale, C.N.R., Via Fosso del Cavaliere,
                I--00133 Roma, Italy
     }

   \date{Received; accepted}


\abstract{
We present and discuss BeppoSAX observations of  Mrk~841 and Mrk~335,
two Seyfert 1 galaxies 
in which previous observations have established the presence
of soft excesses. We confirm the soft excess in both sources, even if
for Mrk~841 a warm absorber provides a fit almost as good as the one
with a true excess. \\ 
As far as the hard X--ray continuum is concerned, a Comptonization
model provides a fit as good as a power law and a physically sound
solution for Mrk~841. For Mrk~335, the Comptonization model gives a 
result which is 
somewhat better on statistical ground, but rather problematic
on physical ground. \\
The most interesting results regard the reprocessing components. For
Mrk~841 we find a very large reflection continuum but an almost normal
iron line equivalent width even if, within the errors, a solution in 
which both components are a factor $\sim$2 larger than expected for an
accretion disc is still marginally acceptable. If this is the 
case, an anisotropy
of the primary emission seems the best explanation. On  the contrary, in
Mrk~335 we find a very large iron line EW but a reflection component
not accordingly large. In this case, the best solution seems to be
in terms of reflection from an ionized disc. 
}   

   \maketitle

\section{Introduction}

Soft excesses, i.e. emission below $\sim$1 keV in excess of the power law 
component dominating the X--ray spectrum above that energy, were 
discovered in Seyfert 1 galaxies 
by EXOSAT (e.g. Arnaud et al. 1985; Turner \&
Pounds 1989) and later confirmed by ROSAT (e.g. Walter \& Fink 1993;
Walter et al. 1994; Piro et al. 1997). 
In the first case, however, it is possible that the
excesses were somewhat overestimated because the Compton reflection
component had not yet been discovered at that time, 
and the best--fit hard X-ray spectra
were therefore found flatter than real. In the second case, the limited ROSAT
bandwidth implied that data from another satellite (usually either 
GINGA or ASCA), not always simultaneous, had to be employed,
with related uncertainties. Moreover, it is possible that most of the 
published ROSAT power law indices are steeper than real due to calibration
problems (Fiore et al. 1994; Iwasawa et al., 1999; Mineo et al. 2000), 
again overestimating (if not creating altogether) soft
excesses, when compared with higher energy spectra. ASCA energy band
goes down only to 0.6 keV (and only with the SIS instruments), not enough to
study the majority of soft excesses in Seyfert 1s, which were found to
be relevant below that energy.

Despite the lower sensitivity with respect to both ROSAT and ASCA in
the overlapping energy ranges, BeppoSAX (Boella et al. 1997), 
thanks to the broad energy covering (0.1-200 keV), may overcome some of the
problems affecting previous studies on soft excesses. While confirming the
commonness of soft excesses in both Narrow Line Seyfert 1 galaxies (Comastri
et al. 2000, and references therein) and in high luminosity PG quasars 
(Mineo et al., 2000), up to now BeppoSAX observations seem indicate that
soft excesses are the exception rather than
the rule in classical, broad line Seyfert 1 galaxies (Matt 2000 and references
therein). We therefore decided to study with BeppoSAX 
two Seyfert 1s in which the claim
of the presence of soft excesses have been among the strongest in the past.

Mrk~841 is indeed the first source in which the soft excess was discovered
(Arnaud et al. 1985).  Since then, Mrk~841 has been extensively studied,
remaining one of the best case for this component
(George et al. 1993; Nandra et al. 1995). In particular, it is one of the 
three (out of 24)
sources in the ASCA sample of Reynolds (1997) in which ``soft excess
is clearly seen in the data" (the other two sources being Mrk~335, see
below, and NGC~4051).
The source is interesting also in other respects: the intrinsic power
law index varies strongly between different observations
(which, by the way, may confuse the soft excess estimate if non--simultaneous 
observations in different bands are compared); a fairly
strong fluorescent iron K$\alpha$ 
line (EW$\sim$450 eV) was detected by GINGA (George et al. 1993); in one GINGA
observation the source was apparently reflection--dominated (George et al.
1993), similarly
to what happened recently to NGC~4051 (Guainazzi et al. 1998). All
these facts make Mrk~841 a source worth investigating even independently of
the soft excess issue.

Mrk~335 is another source in the Reynolds (1997) sample in which the soft
excess is clearly present, confirming previous findings 
by EXOSAT (Turner \& Pounds 1989) and BBXRT (Turner et al. 1993a).
Similarly to Mrk~841, no clear evidence for warm absorption is present
in ASCA data (Reynolds 1997; George et al. 1998), but Turner et al.
(1993b) previously found evidence for a warm absorber
in the ROSAT data. Evidence for such an absorber was also found by 
Orr (2000) in the same BeppoSAX data we are discussing in this paper.
Both Reynolds (1997) and Orr (2000) detected a strong iron K$\alpha$ line,
with EW$\sim$560 and 230 eV, respectively. It should be noted that
Mrk~335 is often classified as a NLSy1, but has a relatively high H$\beta$
FWHM (1640 km/s, Wang et al. 1996,  close to the conventional boundary 
for this class of objects of 2000 km/s), and a rather normal power law index
(see also below), and may therefore be considered a borderline object. 

\section{Observations and data reduction}

\begin{figure}
   \centering
   \includegraphics[width=12.cm]{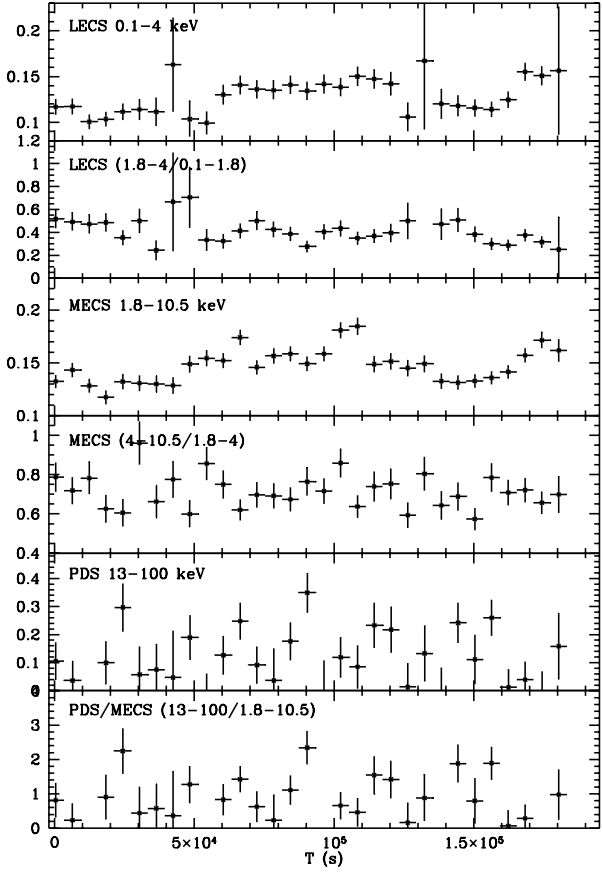}
\caption{Mrk~841. From top to bottom: LECS light curve and hardness ratio; MECS
light curve and hardness ratio; PDS light curve (background subtracted)
and PDS/MECS hardness ratio.
}
\label{lc_841}
\end{figure}

\begin{figure}
   \centering
   \includegraphics[width=12.cm]{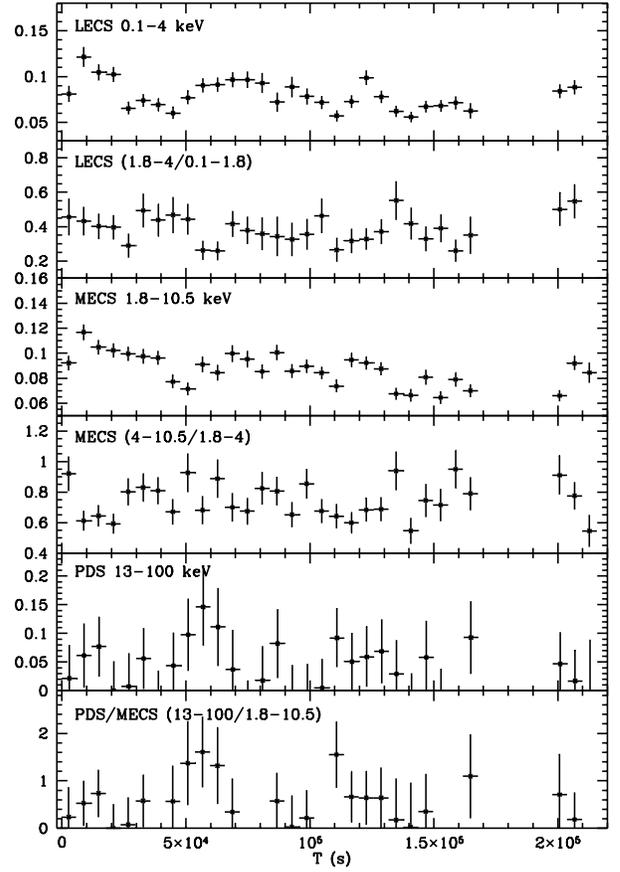}
\caption{Mrk~335. From top to bottom: LECS light curve and hardness ratio; MECS
light curve and hardness ratio; PDS light curve (background subtracted)
and PDS/MECS hardness ratio.
}
\label{lc_335}
\end{figure}

For both sources, 
the observing dates, exposure times, count rates (for the LECS, MECS
and the PDS instruments, the latter referring to two units, as
at any time only half detector is on source), 
fluxes and luminosities are summarized in Table~1. While
the Mrk~841 data are presented here for the first time, results from
the BeppoSAX observation of Mrk~335 have already been presented, 
but not discussed in detail, by Orr (2000). For the
imaging instruments, light curves and spectra have been extracted from
4' (LECS) and 3' (MECS) circles centred on the sources.
Background spectra taken in the same regions
from blank sky observations have been used for subtraction.
Regarding the PDS, background
subtracted spectra automatically generated at the BeppoSAX Scientific Data
Centre (SDC) have been used. Variable Rise Time thresholds have been adopted.

\section{Temporal analysis}

\begin{table*}
\centering
\caption{\label{data}Observation epochs and mean count rates, fluxes
and luminosities ($H_0$=50 km/s/Mpc).}
\vspace{0.05in}
\begin{tabular}{|c|c|cc|cc|cc|c|c|}
\hline
~& ~ & ~ & ~ & ~ & ~ & ~ & ~ & ~ & ~\cr
~ & & \multicolumn{2}{|c|}{LECS} &  
\multicolumn{2}{|c|}{MECS}  &  
\multicolumn{2}{|c|}{PDS} & Flux & Lum. \cr
Target & Start Date & \multicolumn{2}{|c|}{(0.1-4~keV)} & 
\multicolumn{2}{|c|}{(1.8-10.5~keV)} & \multicolumn{2}{|c|}{(13-200~keV)} & 
(2--10~keV) & (2--10~keV) \cr
~& ~ & t$_{\rm exp}$  & CR  & t$_{\rm exp}$ & CR & t$_{\rm exp}$ & 
CR & erg cm$^{-2}$ s$^{-1}$ & erg s$^{-1}$ \cr
~& ~ & (s) & (cts~s$^{-1}$)  & (s) &(cts~s$^{-1}$) & (s) & 
(cts~s$^{-1}$) & ~ & ~ \cr
~& ~ & ~ & & ~ & & ~ & & ~ & ~ \cr
\noalign {\hrule}
\hline
~& ~ & ~ & ~ & ~ & ~ & ~ & ~ & ~ & ~\cr
Mrk~841 & 1999-Jul-30 & 32760 & 0.124
& 87621 & 0.141 & 83008 & 0.12 & 1.3$\times$10$^{-11}$
& 7.3$\times$10$^{43}$ \cr
~& (12h~15m~49s~UT) & ~ & $\pm$0.002 & ~ & $\pm$0.001 & ~ 
& $\pm$0.02 & ~ & ~ \cr
~& ~ & ~ & ~ & ~ & ~ & ~ & ~ & ~ & ~\cr
\hline
~& ~ & ~ & ~ & ~ & ~ & ~ & ~ & ~ & ~\cr
Mrk~335 & 1998-Dec-10 & 40544 & 0.076
& 86455 & 0.091 & 85178 & 0.05 & 0.8$\times$10$^{-11}$
& 2.3$\times$10$^{43}$ \cr
~& (07h~46m~28s~UT) & ~ & $\pm$0.001 & ~ & $\pm$0.001 & ~ & $\pm$0.02 & ~ & ~\cr
~& ~ & ~ & ~ & ~ & ~ & ~ & ~ & ~ & ~\cr
\hline
\end{tabular}
\end{table*}

\subsection{Mrk 841}

The LECS (0.1--4 keV), MECS (1.8-10.5 keV) and PDS (13-100 keV)
light curves of Mrk~841 are shown in Fig.~\ref{lc_841}.
The source varies up to 50\% 
on time--scales of tens of thousand seconds. To search for spectral
variability, we have also plotted the (1.8-4)/(0.1-1.8) LECS, 
the (4-10.5)/(1.8-4) MECS and the PDS/MECS hardness ratios.
There is marginal evidence for moderate and rather erratic spectral 
variability, not correlated with flux variations.

\subsection{Mrk 335}

A similar temporal analysis has been made for Mrk~335
(see Fig.~\ref{lc_335}). During the observation the source varied up to
a factor $\sim$2 in both the LECS and MECS.
There is marginal evidence for spectral variability, but also in this case 
not correlated with flux variations.

\section{Spectral analysis}

For both sources, the spectra have been averaged over the entire observations,
as spectral variability is modest. As usual, 
in the fits the normalization of the LECS and PDS with respect to MECS
has been kept free and fixed to 0.8, respectively.
In the following, all quoted errors 
correspond to the 90 percent confidence level for one interesting parameter
($\Delta\chi^2$=2.7). All spectal fits were performed with the X{\sc spec}
package (version 10).

\subsection{\label{841}Mrk 841}

The data from the three instruments were fitted together.
We first fitted the spectra above 2 keV with a model composed by a power
law with exponential cut--off absorbed by the Galactic $N_{\rm H}$
($2.34\times 10^{20}$ cm$^{-2}$, obtained using the {\sc W3nH} HEASARC tool),
plus a reflection component and an iron K$\alpha$ line. The reflection
component is described by the {\sc pexrav} model 
(Magdziarz \& Zdziarski, 1995), while for the iron line we used either
a narrow gaussian line or a relativistic line ({\sc diskline} model); 
in the latter model we
kept fixed the inner radius at 6$r_{\rm g}$, leaving therefore
as free parameters the outer radius and the inclination angle 
(the latter linked to that of the reflection component).  
We note that a broad gaussian line, besides having no astrophysical meaning
in the context under discussion, has the same number of free parameters
as the relativistic lines, and therefore is not a {\it simpler} model. 
We therefore did not use this line parametrization in our fits. We 
tried to apply the relativistic effects to the reflection component too,
adopting the {\sc refsch} model, but no appreciable difference could be
observed, due to the limited statistics. 

\begin{figure}
   \centering
   \includegraphics[angle=-90,width=9.5cm]{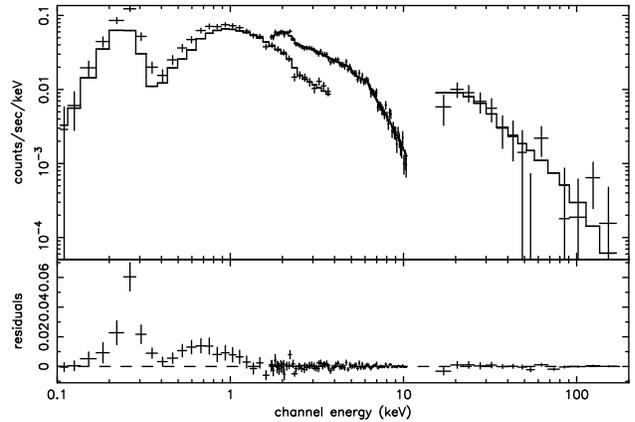}
\caption{\label{extrap_841} BeppoSAX LECS+MECS+PDS data and best fit model 
above 2 keV for Mrk 841, then extrapolated to lower energies to 
illustrate the presence of a soft excess.}
\end{figure}

This model (whatever the line parametrization) provides
a good description of the data between 2 and 200 keV, but when extrapolated
to lower energies (see Fig.~\ref{extrap_841}) it falls significantly short 
of the data,  suggesting the presence of a soft excess. 

When the fit is performed over the whole band, 
this model gives an acceptable fit ($\chi^{2}$=135.4/121 degrees of freedom
(dof)) after steepening the power law index and simultaneously increasing the
reflection component, 
but features in the residuals in the soft part of the spectrum are still
present. We thus checked if the presence of a warm 
absorber is consistent with the data, by adding to 
the baseline model the {\sc absori} component. The addition of this 
component significantly improves the fit
($\Delta\chi^2$(dof) = 13.6(2), corresponding to a probability of
99.8\% according to the F--test). The best fit
parameters are listed in Table~\ref{par841}, first column.
The value of $R$, 4.6, is very high, while the 
equivalent width of the iron line is 210~eV, inconsistent with 
the large value of $R$ found (e.g. Matt et al. 1992). 
There is also a problem with the iron line properties, 
as the value of $r_{\rm out}$ is basically equal to 
$r_{\rm in}$. If we use a simple, narrow  gaussian 
profile ({\sc zgauss} model) instead of the relativistic one for the iron 
line, we get a fit of similar quality ($\chi^{2}$=123.9/120 dof). The value for 
$R$ is also similar ($4.3^{+9.1}_{-1.6}$), but the EW is much lower 
($80^{+40}_{-60}$~eV), making more acute the inconsistency with 
the large value of $R$. In the following fits
we will therefore consider only the relativistic line, keeping in mind
that this choice is based on astrophysical plausibility rather than
statistical arguments. Excluding altogether the line from
the model, we found a still acceptable fit ($\chi^{2}$=129.6/121 dof),
but again this is rather unplausible given the unambiguous presence
of a reflection continuum (in the fit without the line the value of $R$ 
raises to 6.4!).
According to the F-test, the presence of the (relativistic) 
line is significant at the 98.5\% confidence level. 

\begin{figure}
   \centering
   \includegraphics[angle=-90,width=9.5cm]{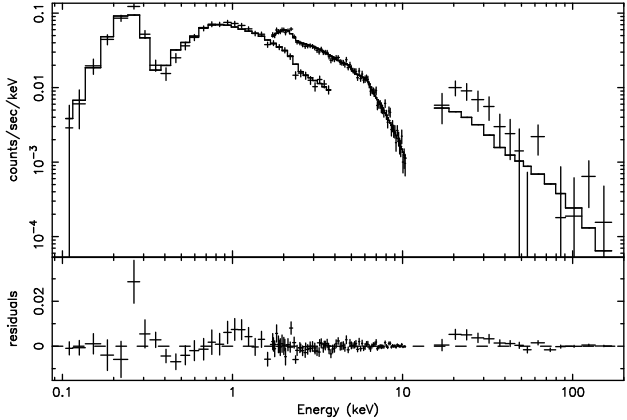}
\caption{\label{badfit_r}BeppoSAX LECS+MECS+PDS data and best fit model 
for Mrk 841, when the reflection component is not included.
An excess of counts in the PDS band is clearly present.}
\end{figure}

Of course, excluding the reflection component leads to a very bad fit,
as illustrated in Fig.~\ref{badfit_r}, obtained by fitting the data with
a simple power law instead of the {\sc pexrav} model.

The next step was to fit the data with the baseline model plus a thermal 
emission ({\sc bb} model), instead of the warm absorber
(second column of Table 2). This results in a 
similar $\chi^{2}$ (120.8/119 dof). The 
equivalent width of the iron line is higher ($260^{+120}_{-80}$ eV), 
thus becoming more consistent with the large value of the reflection component. 
Even if the value for $r_{\rm out}$ is still very low, it is higher than before.
Moreover, $R$ is lower and significantly better 
constrained than in the previous fit, being $3.2^{+1.6}_{-1.0}$. As in the 
previous fit, only a lower limit can be put on the high energy cut--off.

\begin{table*}
\centering
\caption{\label{par841}Best fit parameters for the two models described 
in Sect. \ref{841} for Mrk 841. The values followed by $^{*}$ are kept 
fixed in the fit. }
\vspace{0.05in}
\begin{tabular}{|c|c||c|c|c|}
\hline 
\multicolumn{2}{|c||}{}&
 \textsc{pexrav+absori} model&
 \textsc{pexrav+bb} model & \textsc{Compt.+bb} model
\\
\hline 
\hline 
\textsc{wabs}&
\( N_{\rm H} \)&
\( 2.34\times 10^{20}\, {\rm cm^{-2\, *}} \)&
\( 2.34\times 10^{20}\, {\rm cm^{-2\, *}} \)&
\( 2.34\times 10^{20}\, {\rm cm^{-2\, *}} \)\\
\hline 
\hline 
\textsc{pexrav}&
\( \Gamma  \)&
\( 2.28_{-0.06}^{+0.04} \)&
\( 2.12_{-0.02}^{+0.10} \) &  -\\
\hline 
&
\( R \)&
\( 4.6_{-1.4}^{+0.6} \)&
\( 3.2_{-1.0}^{+1.6} \)&  -\\
\hline 
&
\( \cos i \)&
\( 0.80_{-0.53}^{+0.20} \)&
\( 0.82_{-0.41}^{+0.09} \)&  - \\
\hline 
&
\( E_{\rm c} \)&
$>$150 \,{\rm keV}&
\( >100 \,{\rm keV} \)&  -\\
\hline 
\hline 
\textsc{compt.}&\( T_{\rm soft}  \)& - & - & \( 71_{-8}^{+6} {\rm eV} \) 
\\
\hline 
& \( \cos i \)& - & - & \( 0.8^{*} \) \\
\hline 
& \( \Theta \)& - & - & \( 0.13_{-0.02}^{+0.01} \)    \\
\hline 
& \( \tau \)& - & - & \( 0.80_{-0.06}^{+0.13} \)  \\
\hline
& \( R \)& - & - & \( 3.1_{-0.8}^{+1.6} \)  \\
\hline 
\hline 
\textsc{absori}&
\( \xi  \)&
\( 530_{-260}^{+1000}\, {\rm erg\, cm\, s^{-1}} \)&
- &  -\\
\hline
& \( N_{\rm H} \)&
\( \left( 1.1_{-0.5}^{+1.6}\right) \times 10^{22}\, {\rm cm^{-2}} \)&
- &  - \\
\hline 
\hline 
\textsc{bb}&
\( kT \)&
-&
\( 95_{-10}^{+10}\, {\rm eV} \) &
\( 182_{-45}^{+56}\, {\rm eV} \) 
\\
\hline
\hline
\textsc{zdisk}&
\( E \)&
\( 6.4\, {\rm keV^{*}} \)&
\( 6.4\, {\rm keV^{*}} \)&
\( 6.4\, {\rm keV^{*}} \)\\
\hline 
&
\( EW \)&
\( 210_{-140}^{+120}\, {\rm eV} \)&
\( 260_{-80}^{+120}\, {\rm eV} \)&
\( 390_{-140}^{+150}\,{\rm  eV} \)
\\
\hline 
&
\( r_{\rm out} \)&
\( 6.01_{-0}^{+4.4}\,r_{\rm g} \)&
\( 9.0_{-3.0}^{+6.3}\,r_{\rm g} \)&
\( 6.01_{-0}^{+4.8}\,r_{\rm g} \) 
\\
\hline 
\hline 
\multicolumn{2}{|c||}{ \( \chi ^{2}/dof \) }&
121.8/119&
120.8/119&
118.8/120
\\
\hline 
\hline 
\multicolumn{2}{|c||}{N.H.P.}&
0.41&
0.44&
0.51
\\
\hline 
\end{tabular}
\end{table*}

\begin{figure}
   \centering
   \includegraphics[angle=-90,width=9.5cm]{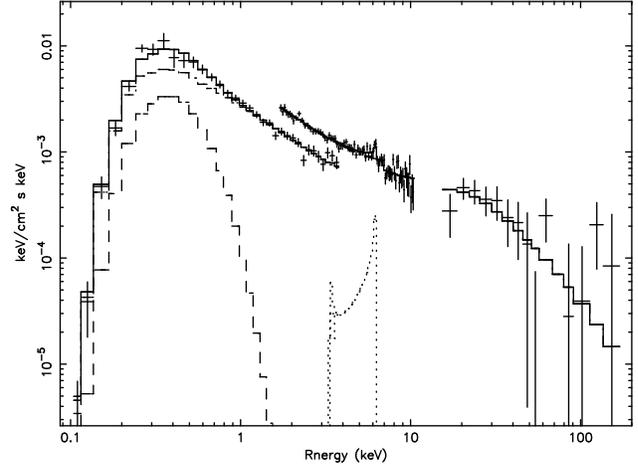}
\caption{\label{841bb}BeppoSAX LECS+MECS+PDS unfolded spectrum 
 and best fit model (second column of Table 2) for Mrk 841.}
\end{figure}

Since the value of the solid angle of the reflection component is still 
very high, we checked whether this could be entirely ascribed to an excess of
counts in the PDS. Thus we fitted 
the same model with LECS and MECS data only, without the PDS. The resulting 
$\chi^{2}$ is 108.9/105 dof: the value of $R$ is still very high ($\simeq3$),
therefore excluding that it may arise from a miscalibration between
the MECS and the PDS instruments, or to contaminating sources in the PDS
field of view.

We also tried to model the soft excess with a power law instead of a 
black body. The fit is significantly worse ($\chi^{2}$=131.5/119 dof), 
even adding a warm absorber ($\chi^{2}$=125.8/117 dof), but in the last
case a lower value for $R$ (2.7) and a more consistent iron line EW (250 eV)
is found. 

On the contrary, the fit with the bremsstrahlung emission provides a fit
almost as good ($\chi^{2}$=123.4/119 dof) as with the black body.
The value of $R$, 2.3, is lower and consistent, within the errors,
with the value of the iron line EW, 280 eV.

We next tried to fit the primary continuum with a more physical model
than a simple power law; we used the Comptonization model of 
Haardt \& Maraschi (1993) (third column of Table 2). 
The corresponding XSPEC code is in the form of a grid of spectra;
the fitting parameters are the electron temperature in units of $mc^2$,
$\Theta$,  the optical depth $\tau$, and the temperature of the soft
 black body emission, $T_{\rm soft}$; the parameters range
from 0.1 to 1 ($\Theta$), 0.05 to 1 ($\tau$) and 5 to 100 eV ($\Theta$). 
The cosine of the inclination angle has been fixed to 0.8 (as suggested
by the {\sc pexrav} plus {\sc diskline} fit), in order to reduce the 
number of free parameters. The Compton reflection component is also included in 
the model. The fit is satisfactory 
($\chi^{2}$=118.8/120 dof), but the problem of the large value of $R$ 
is not solved.
The line EW is about 390 eV (obtained with the diskline model).
 The parameters of the continuum are $\tau$=0.8 
and $\Theta$=0.13. It is worth noting that the soft emission present in the 
Comptonization model is not sufficient, and a
further black body component is also required, suggesting either a multicolor
disc emission or an altogether different component. 

If a bremsstrahlung is used instead of a black body, the fit is significantly 
worse ($\chi^{2}$=135.0/120 dof). A power law, instead, provides a fit of
comparable quality ($\chi^{2}$=121.0/120 dof), 
but the problem of the large value of $R$ is made even more 
serious. Finally, a warm absorber, instead of a true soft excess, 
gives a significantly worse fit ($\chi^{2}$=138.6/120 dof).

\subsection{\label{335}Mrk 335}

\begin{figure}
   \centering
   \includegraphics[angle=-90,width=9.5cm]{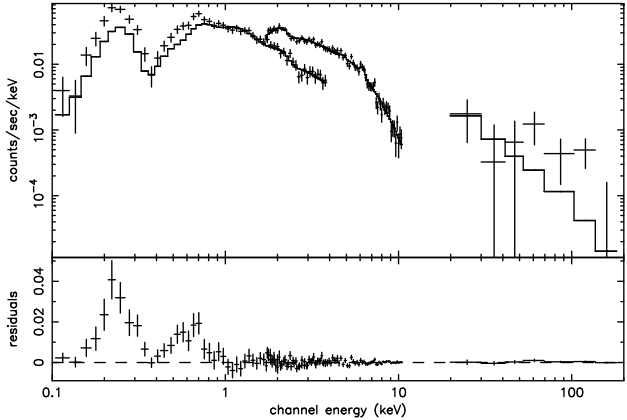}
\caption{\label{extrap_335} BeppoSAX LECS+MECS+PDS data and best fit model 
above 2 keV for Mrk 335, then extrapolated to lower energies to 
illustrate the presence of a soft excess.}
\end{figure}

\begin{figure}
   \centering
   \includegraphics[angle=-90,width=9.5cm]{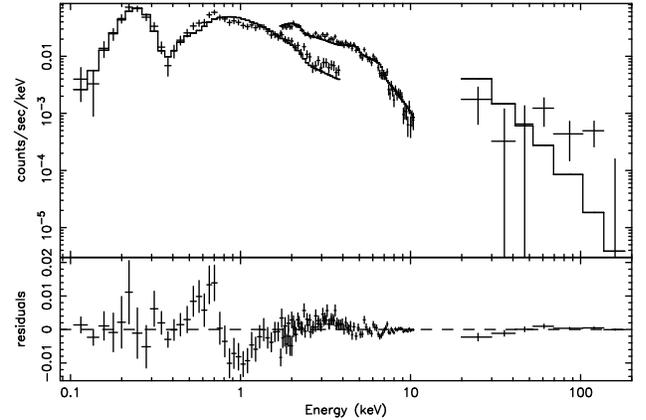}
\caption{\label{335soft}Data and best fit model for
 Mrk 335, when the baseline model is adopted. 
The fit is clearly inadequate below $\sim$2 keV, due both to the
warm absorber and a soft excess (see text for details).}
\end{figure}

We firstly proceeded as for Mrk~841, and fitted the data above 2 keV with
the same baseline model, with a Galactic
column density $N_{\rm H}=4.0\times 10^{20}$ cm$^{-2}$.
Given the poor statistics in the PDS, we did not include the
high energy cut--off. The extrapolation to lower energies is shown
in Fig.~\ref{extrap_335}: the soft excess is even more clear than for
Mrk~841. In fact, fitting this model to the whole band, 
we obtain a totally unacceptable fit, with 
$\chi^{2}$=286.4/128 dof (see Fig.~\ref{335soft}). The inclusion of a warm absorber strongly 
improves the fit ($\chi^{2}$=151.4/126 dof), but still a power law is needed to 
model the soft excess ($\chi^{2}$=134.2/124 dof).  The solid angle of the 
reflection component is rather low ($\sim$0.6) and ill constrained, 
while the equivalent 
width of the iron line is rather high, being $\sim$300 eV. If a narrow
gaussian line is used instead of the relativistic one, the fit is
worse ($\chi^{2}$=139.3/125 dof)

A significant improvement in the fit  ($\Delta\chi^{2}\simeq8.2$,
corresponding to a probability of about 99.4\% according to the F--test) is 
obtained adding a narrow gaussian line to the relativistic one. The 
new model (see the first column of Table \ref{par335})
has parameters not very different from the previous one, but the total 
equivalent width of the iron line is very large, the sum of the 
two components being 270+390=660 eV.

If a black body or a bremsstrahlung 
is used instead of the power law,  rather worse
fits are found ($\chi^{2}$=130.7/123 dof for the black body, 
$\chi^{2}$=133.2/123 dof for the bremsstrahlung). 

Recently, Ballantyne et al. (2001) have found that a good fit to the ASCA
spectrum of Mrk~335 is achieved with the ionized disc model (Ross \&
Fabian 1993). This could naturally explain the large equivalent width of
the iron line (Matt et al. 1993, 1996). We therefore substituted the model
{\sc pexriv}, which allows for ionization of the reflecting matter, to
{\sc pexrav}, which is for neutral matter alone. 
We also fixed the energy of the 
iron line in the {\sc diskline} model either to 6.7 (He--like iron) or
to 6.97 (H--like iron), and excluded from the fit the narrow line. 
The results are inconclusive: the $\chi^2$ is not improved
(126.1/123 dof for the H--like iron, and 126.8/123 dof  for
the He--like iron), and the reflection component in both cases goes to zero
(but remaining ill--constrained). The equivalent width of the iron line
is 690 eV for the H--like iron, and 660 eV for the He--like iron.
Very similar results are obtained
using the Ross \& Fabian (1993)\footnote{available at 
{\sc http://heasarc.gsfc.nasa.gov/\-docs/\-xanadu/\-xspec/\-models/\-iondisc.html} }
model. We therefore conclude that the ionized disc model is neither proved
nor disproved by the data, remaining the more plausible on physical ground.

\begin{table*}
\centering
\caption{\label{par335}Best fit parameters for the two models described in 
Sect. \ref{335} for Mrk 335. The values followed by $^{*}$ are kept fixed 
in the fit. }
\vspace{0.05in}
\begin{tabular}{|c|c||c|c|}
\hline 
\multicolumn{2}{|c||}{}&
 \textsc{pexrav+pow}  & 
\textsc{Compt.+bb} \\
\hline 
\hline 
\textsc{wabs}&
\( N_{\rm H} \)&
\( 4.0\times 10^{20}\, {\rm cm^{-2\, *}} \)&
\( 4.0\times 10^{20}\, {\rm cm^{-2\, *}} \)\\
\hline 
\hline 
\textsc{pexrav}&
\( \Gamma  \)&
\( 2.04_{-0.03}^{+0.17} \) &  - \\
\hline 
&
\( R \)&
\( 0.58_{-0.58}^{+1.47} \) &  - \\
\hline 
&
\( \cos i \)&
\( 0.90_{-0.20}^{+0.05} \) &  - \\
\hline 
&
\( E_{\rm c} \)&
\( \infty^{*} \) &  - \\
\hline 
\hline 
\textsc{compt.} & \( T_{\rm soft}  \) & -  
& \( 98_{-4}^{+2} {\rm eV} \) \\
\hline
& \( \cos i \)& - & \( 0.8^{*} \) \\
\hline
& \( \Theta \)& -  & \( 0.14_{-0.03}^{+0.03} \) \\
\hline
& \( \tau \)& - & \( 0.62_{-0.09}^{+0.16} \) \\
\hline
& \( R \)& - & \( 0_{-0}^{+1.6} \)  \\
\hline
\hline
\textsc{absori}&
\( \xi  \)&
\( 510_{-40}^{+60}\, {\rm erg\, cm\, s^{-1}} \) & 
\( 890_{-740}^{+1060}\, {\rm erg\, cm\, s^{-1}} \) 
 \\
\hline 
&
\( N_{\rm H} \)&
\( \left( 4.2_{-0.3}^{+0.7}\right) \times 10^{22}\, {\rm cm^{-2}} \)&
\( \left( 2.1_{-1.7}^{+2.2}\right) \times 10^{22}\, {\rm cm^{-2}} \)
\\
\hline 
\hline 
\textsc{powerlaw}&
\( \Gamma  \)&
\( 2.68_{-0.04}^{+0.08} \)&  - \\
\hline 
\hline 
\textsc{bb}&
\( $kT$  \)& - &
\( 29_{-7}^{+7} {\rm eV} \) \\
\hline 
\hline 
\textsc{zgauss}&
\( E \)&
\( 6.4\, {\rm keV^{*}} \)&  -
\\
\hline 
&
\( EW \)&
\( 270_{-70}^{+65}\, {\rm eV} \)&  - 
\\
\hline 
\textsc{zdiskline}&
\( E \)&
\( 6.4\, {\rm keV^{*}} \)&
\( 6.4\, {\rm keV^{*}} \)\\
\hline 
&
\( r_{\rm out} \)&
\( 6.01_{-0}^{+3.3}\, r_{\rm g} \)&
\( 6.02_{-0.01}^{+1.62}\, r_{\rm g} \)
\\
\hline 
&
\( EW \)&
\( 390_{-89}^{+86}\, {\rm eV} \)&
\( 760_{-220}^{+230}\, {\rm eV} \)
\\
\hline 
\hline 
\multicolumn{2}{|c||}{ \( \chi ^{2}/dof \) }&
126.0/123&
121.4/123
\\
\hline 
\hline 
\multicolumn{2}{|c||}{ N.H.P.}&
0.41&
0.53
\\
\hline 
\end{tabular}
\end{table*}

We then fitted the spectrum with the Comptonization model described in
the previous section, using a power law, a black body or a 
bremsstrahlung model for the soft excess. In all cases, the fit is 
slightly better than that with a cut--offed power law. The best fit
is found for a black body (121.4/123 dof, second column of Table~3), while for
the power law and the bremsstrahlung we found a reduced $\chi^2$ of
124.5/123 dof and 123.2/123 dof, respectively.
None of these fits require a narrow iron line 
besides the relativistic one. 

As far as the reprocessing components are concerned, we still found a very
large iron line, and a Compton reflection component consistent with zero.

\begin{figure}
   \centering
   \includegraphics[angle=-90,width=9.5cm]{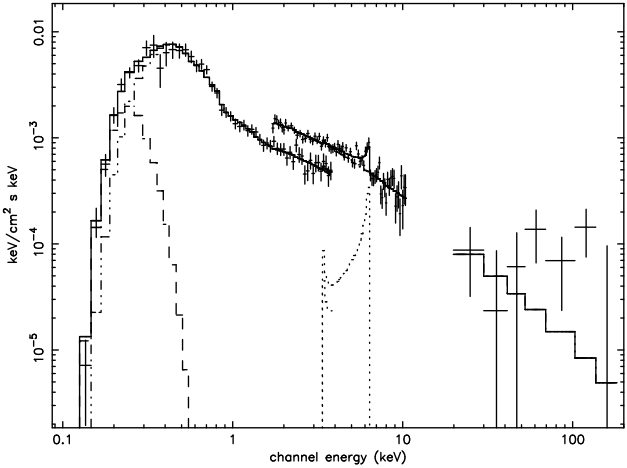}
\caption{\label{335fit}BeppoSAX LECS+MECS+PDS unfolded spectrum
 and best fit model (second column of Table~3) for  Mrk 335.}
\end{figure}

\section{Discussion and Conclusions}

We analysed the BeppoSAX data of Mrk~841 and Mrk~335, two Seyfert 1s in 
which previous observations indicated the presence of a soft excess.
We indeed confirm the presence of this component, even if for  Mrk~841
the fit with the warm absorber is almost as good as that with a true soft
excess. The quality of the data
are not, unfortunately, good enough to choose between different 
parameterizations of the soft emission. In fact, for Mrk~841 we found that
a power law is a significantly poorer description of the soft excess, but
a black body and a bremsstrahlung are equally acceptable (even if the former
is more plausible if the soft excess is indeed associated with the accretion
disc), at least as long as a power law is used for the hard X--ray component. 
 For Mrk~335, the presence of the soft excess is clear. If a simple
power law is used for the hard component, then a power law is also to be
preferred for the soft excess. If, instead, the hard X--ray emission
is described by the Comptonization model of Haardt \& Maraschi (1993), 
the black body and the bremsstrahlung provide a better fit. 
To quantify the relative importance of the soft excess, in Table 4
the bolometric and 0.1-0.5 keV fluxes in the black body component, for
both the {\sc pexrav} and Comptonization models, are given for both
sources. Comparing them with the 2-10 keV fluxes given in Table 1, 
it is clear that the soft excess is more important in Mrk~335, not
surprisingly as it is a NLS1. Rather unexpectedly, for Mrk~335 the largest 
fluxes are obtained in the Comptonization model, despite the presence
of a further thermal component in the model.

Comparing our results with previous observations, we note that the soft
excess in Mrk~841 is significantly weaker than that found by Arnaud et al.
(1985), who however found a much flatter spectral index ($\Gamma$=1.6)
for the hard component. Because at that time the spectral complexities
of Seyfert 1s have not yet been discovered, it is possible that they
overestimated the soft excess (see the Introduction). Our results are
instead remarkably similar to those found by Nandra et al. (1995) analysing
ROSAT data (compare model B of their Table 1 with the second column of 
our Table 2 and with Table 4). 

As far as the soft excess in Mrk~335 is concerned, comparing our results
with those of Reynolds (1997, see his Table 5), 
based on ASCA data, we note that the 
black body bolometric luminosity in our observation is a factor 2.7 higher
than in the ASCA observation, while the 
2-10 keV luminosities are very similar (with the photon indices being also
very similar). This finding rules out models in which the soft excess is
entirely due to reprocessing of the hard X--ray component, because in such
a model the two components should vary together.

\begin{table}
\centering
\caption{\label{fluxes_bb}Fluxes and luminosities
in the black body component.}
\vspace{0.05in}
\begin{tabular}{|c|cc|cc|}
\hline
~& ~ & ~ & ~ & ~\cr
~ & \multicolumn{2}{|c|}{\sc pexrav+bb} &  
\multicolumn{2}{|c|}{\sc Compt+bb} \cr
~ & \multicolumn{2}{|c|}{Flux/[Lum.]} & \multicolumn{2}{|c|}{Flux[Lum.]}\cr
~ & \multicolumn{2}{|c|}{(10$^{-11}$~erg~cm$^{-2}$~s$^{-1}$)} &
\multicolumn{2}{|c|}{(10$^{-11}$~erg~cm$^{-2}$~s$^{-1}$)} \cr
~ & \multicolumn{2}{|c|}{[(10$^{43}$~erg~s$^{-1}$)]} &
\multicolumn{2}{|c|}{[(10$^{43}$~erg~s$^{-1}$)]} \cr
Source & Bol. & 0.1-0.5~keV  & Bol. & 0.1-0.5~keV \cr
\noalign {\hrule}
\hline
~& ~ & ~ & ~ & ~\cr
Mrk~841 & 0.69 & 0.51 & 0.36 & 0.12 \cr
~ & [3.9] & [2.9] & [2.0] & [0.7] \cr
~& ~ & ~ & ~ & ~\cr
\hline
~& ~ & ~ & ~ & ~\cr
Mrk~335 & 1.2 & 1.0 & 5.6 & 2.9 \cr
~ & [3.5] & [2.9] & [16.1] & [10.9] \cr
~& ~ & ~ & ~ & ~ \cr
\hline
\end{tabular}
\end{table}

The Comptonization model gives a slightly better fit for  Mrk~335, and a comparable
one for Mrk~841. While the quality of the spectra, especially
at high energies, is not good enough to allow definitive conclusions, it
may be instructive to compare the present results with those obtained,
adopting the very same model, by Petrucci et al (2001, P01) on a sample of 
Seyfert 1s observed by BeppoSAX. For both sources the best fit value 
for the temperature of the plasma is about 70 keV, significantly lower than
any value in the P01 sample. Looking at their Fig.~2, we
can see that the values we find of $\Theta$ and $\tau$ are in between
the theoretical curves for a plane--parallel (`slab') geometry and for
an hemispherical one (in this respect, in agreement with most
of the sources analyzed by P01; note also that, for Mrk~335, a pure
plane--parallel
geometry is ruled out by the comparison of BeppoSAX and ASCA
results, see above). The need for a further black body
component may suggest that the correct geometry is the patchy one,
as envisaged by Haardt et al. (1994), at least for Mrk~841. In the
case of Mrk~335, instead, the fact that the temperature of the
extra black body is greater than that of the Comptonized one makes
this solution hardly tenable.

For both sources the most puzzling results 
concern the reprocessed components. In Mrk~841 we found a very large value
for the solid angle, $R$, of the Compton reflection component, while 
the EW of the iron line, supposed to originate in a relativistic disc,
is larger than usual but not so extreme, at least as far as the face
values are considered. Actually, within the errors a solution in which
both components are about a factor 2 larger than in the $\Omega=2\pi$ case
is possible (in agreement with previous GINGA observations,
in which very large values for both the iron line EW and the amount
of reflection continuum were found, George et al. 1993). 
This may be explained either by an anisotropy of the 
illuminating radiation (see e.g. Ghisellini et al. 1991, where it is shown
that even larger anisotropies may be achieved), 
or by a delayed response of the reprocessing
components to variations of the primary continuum (but the latter possibility 
implies a distance of the reprocessing matter greater 
than that of the inner accretion disc,
in disagreement with the hypothesis of a relativistic iron line necessary
to have a sufficent EW to match the large amount of reflection). 
Interestingly, the large value of $R$ seems to follow the anticorrelation 
between $R$ and $\Theta$ found by P01, as well as that the correlation
between $R$ and $\Gamma$ found by Zdziarski et al. (1999).

In Mrk~335, on the contrary, we found a very large iron line (in agreement 
with previous findings) but 
small or moderate Compton reflection component (the upper limit being
between 1 and 2, depending on the model for the continuum). 
A possible explanation
could be a larger--than--solar iron abundance, but the values required are 
rather high, i.e. around 10 (Matt et al. 1997). A more viable solution 
is in terms of an ionized disc; if the iron is mainly in He-- and H--like
ions, an equivalent width several times that for neutral iron can be
obtained, at least in the constant density solution (Ross \& Fabian 1993;
Matt et al. 1993, 1996). It is worth noting that Mrk~335 is usually classified
as a Narrow Line Seyfert 1 galaxy, a class of sources in which significant
ionization of the accretion disc are often observed (see e.g. Comastri 2000
and references therein),
not surprisingly as these sources are commonly believed to accrete at high
rates.

\begin{acknowledgements}
The BeppoSAX satellite is a joint Italian--Dutch program. We 
thank the BeppoSAX Scientific Data Centre for assistance,
M. Guainazzi for helpful discussions, and the anonymous referee for
comments that helped improving the clarity of the paper. 

This research has made use of the NASA/IPAC Extragalactic Database (NED)
which is operated by the Jet Propulsion Laboratory, California Institute of
Technology, under contract with the National Aeronautics and Space
Administration. 

This work was supported by the Italian Space Agency, and by the Ministry
for University and Research (MURST) under grants {\sc COFIN98--02--32} and
{\sc COFIN00--02--36}.
\end{acknowledgements}

\end{document}